# Kinematical and Ellipsoidal Properties of the Inner-Halo Hot Subdwarfs Observed in Gaia DR3 and LAMOST DR7

**W. H. Elsanhoury[1,2]**

[1] Department of Physics, College of Science, Northern Border University, Arar, Saudi Arabia.
Email: elsanhoury@nbu.edu.sa

[2] Department of Astronomy, National Research Institute of Astronomy and Geophysics (NRIAG), 11421, Helwan, Cairo, Egypt. Email: elsanhoury@nriag.sci.eg

**Abstract**

Here, we report the kinematical parameters of inner-halo hot subdwarfs located within (d ≤ 15 kpc) at high Galactic latitudes ($b° ≥ 20$). The study included three program stars for one of the extreme He-rich groups (*e*He-1) with eccentricity (e = 0.65) and the z-component of the angular momentum ($J_z$ = 4288.66 kpc km s$^{-1}$), the inner halo program I with 121 points ($T_{eff} ≥ 24{,}000$) and their subsections, i.e. inner halo program II (sdB; 79 points) with ($40{,}000 ≥ T_{eff} ≥ 24{,}000$) and inner halo program III (sdO; 42 points) with ($80{,}000 ≥ T_{eff} ≥ 40{,}000$). First, we calculated the spatial velocities ($\overline{U}, \overline{V}, \overline{W}$; km s$^{-1}$) along the Galactic coordinates (i.e., 25.73 ± 5.07, 28.79 ± 5.37, −14.51 ± 3.81) and their dispersion velocities ($\sigma_1, \sigma_2, \sigma_3$; km s$^{-1}$) = (161.94 ± 12.73, 140.31 ± 11.85, 101.57 ± 10.08) and subsequently their subsections sdB and sdO. Second, we calculated the vertex longitudes ($l_2$) and the Solar motion ($S_\odot = 41.24 \pm 6.42$ km s$^{-1}$) as well as their subsections. Finally, based on the kinematic relation of the ratio ($\sigma_2/\sigma_1$) and our previously computed numerical value of the angular rotation rate ($|A - B| = 26.07 \pm 5.10$; km s$^{-1}$kpc$^{-1}$), we obtained the average Oort's constants as (A & B; km s$^{-1}$kpc$^{-1}$) = (9.38 ± 0.33, −16.69 ± 0.25).

## 1. Introduction

Subluminous stars with luminosity class VI are called subdwarf (sd) stars, which were discovered by Greenstein (1954). They are fainter than main-sequence (MS) stars of similar spectral type and their absolute magnitude criteria in the G-band ($-5 < M_G < 7$) (Culpan et al. 2022). Subdwarfs appear to lie below MS stars on a Hertzsprung–Russell Diagram (HRD) (Heber 2009), they are located at the extreme horizontal branch (EHB) (Heber 2009; 2016), also known as the blueward extension of the horizontal branch (HB).

Hot sd stars are common outcomes of stellar evolution and are essential for understanding the characteristics of ancient stellar populations. Hot sd stars are in the late stages of stellar evolution when a red giant star loses its outer hydrogen layers before the core begins to fuse helium. They have ($T_{eff} ≥ 20{,}000$ K), which is higher than that of MS stars with similar temperatures, although they exhibit lower luminosity. As hot sd stars make up a significant portion of the dim blue star population (i.e., evolved He core burning stars) at high Galactic latitudes compared with hot MS stars of identical spectral types, they are significantly smaller, have very thin hydrogen-dominated atmospheres (nHe/nH ≤ 0.01), masses of ~0.5 $M_⊙$ (Heber 1986), and radii (≃ 0.10 to 0.30 $R_⊙$) (Geier 2015).

Hot sd stars are like typical O/B type stars in their spectral features (with a lower order of O/B magnitudes) and were previously categorized as B-type subdwarf (sdB) and O-type subdwarf (sdO) stars based on their spectral appearance (Drilling et al. 2013). Subdwarf B (sdB) with (40,000 ≥ $T_{eff}$ ≥ 20,000 K) (Martin et al. 2017), which is composed of core helium-burning stars located near the blue end of the HB of the HRD with much broader Balmer lines. sdB exhibits strong and broad Balmer lines but weak (faint or absent) He I and He II lines (Sargent & Searle 1968). Subdwarf O (sdO) with (80,000 ≥ $T_{eff}$ ≥ 40,000 K) (Martin et al. 2017). It is a diverse mixture of post-red-giant branch, post-HB, and postasymptotic giant-branch stars. sdO stars exhibit weak He II lines but strong Balmer lines (Sargent & Searle 1968; Heber 2009). In contrast, subdwarf OB (sdOB) stars represent a transition between the O and B types (Moehler et al. 1990; Heber 2009).

sbB and sdO are found in all Galactic stellar populations. For several reasons, hot sd have been thoroughly investigated. They are common enough to explain the observed UV excess in early-type Galaxies (Heber 2009 and O'Connell 1999). Asteroseismology has become heavily reliant on pulsating sdB stars (Charpinet et al. 2010). Chemical anomalies exist in the atmospheres of sdO/Bs (Geier 2013). Hot subdwarf stars are also relevant for cosmology, as some of them are candidate progenitors of Type Ia supernova (SN Ia) progenitors are sdB stars in near binaries (Maxted et al. 2000). Furthermore, substellar partners such as planets orbiting hot and brown dwarfs have been discovered (Silvotti et al. 2007).

Inner-halo hot sd (d ≤ 15 kpc) maybe utilized to probe the formative phases of our Galaxy (Carney & Latham 1986). The Galactic halo has been broken down into inner and outer halos using astrophysical models (McCarthy et al. 2012). Stars that developed inside the Galaxy made up most of the stars in the inner halo, whereas stars accreted during merger events made up the stars in the outer halo. Blue Horizontal Branch stars (BHB) are ideal tracers for studying the profile of the outer stellar halo ($R_{GC}$ > 20 kpc) and determine the profile of the outer smooth stellar halo up to a Galactocentric radius of ~220 kpc. They are present in old stellar populations, and their absolute magnitude is roughly constant and bright ~ 0.5 mag (Deason, Belokurov & Evans 2011), meaning that they can be identified even at very large distances (> 100 kpc). Gilmore and Reid (1983) reported that the disk of the Milky Way can be split into thin and thick disks. The kinematics, metallicity, and spatial distribution of these components are different (Yan et al. 2019).

The remainder of this paper is presented as follows. Section 2 provides a brief explanation of the data used. Section 3 deals with semidefinite programming (SDP) and velocity ellipsoid parameters (VEPs). Section 4 discusses the structural properties and derivation of the Galactic rotation kinematics. Section 5 presents the results of the study. Finally, Section 6 presents the discussion and conclusions.

## 2. Data and Sample Selection

LAMOST (Large Sky Area Multi-Object Fiber Spectroscopic Telescope), also named the "Guo Shou Jing" Telescope (Cui et al. 2012) is a 4-m Schmidt spectroscopic survey telescope specifically developed to study four thousand targets per exposure in a field of view roughly 5° in diameter (Cui et al. 2012; Luo et al. 2012, 2014). LAMOST spectra have a resolution of ∼1800 with a wavelength range of 3800–9100 Å. In March 2020[1], 10,608,416 spectra in DR7 (Lei et al. 2020) were made accessible by LAMOST.

Global Astrometric Interferometer for Astrophysics (Gaia) (Gaia Collaboration et al. 2016) is the mission of the European Space Agency (ESA) that aims to create the most extensive and accurate 3D map of our Galaxy through an unmatched survey of 1% of the Galaxy's 100 billion stars with a precision of microarcsecond (μas). Gaia DR2, which was released on April 25, 2018, offers three broadband magnitudes (G, $G_{BP}$, and $G_{RP}$) and high-precision locations (α, δ), proper motions $\left(\mu_\alpha^{*[2]}, \mu_\delta; \text{mas yr}^{-1}\right)$, and parallaxes ($\overline{\omega}$; mas) for about 1.3 billion stars brighter than G = 21 mag (Gaia Collaboration et al. 2018). After its release on December 3, 2020, Gaia EDR3 (Gaia Collaboration et al. 2021) substantially increased the accuracy and precision of broadband photometry and astrometry compared with Gaia DR2. Gaia EDR3 (Gaia Collaboration et al. 2021) contains astrometry and photometry for 1.8 billion sources brighter than G = 21 mag. For 1.5 billion of these sources, parallaxes ($\overline{\omega}$), proper motions ($\mu_\alpha^*, \mu_\delta$), and the $G_{BP}$–$G_{RP}$ colors are available. Bailer-Jones et al. (2021) also provided a Gaia EDR3 distance catalog based on Gaia EDR3. Therefore, where necessary, unreliable distances were replaced with the estimated values from the Gaia DR3 distance catalog (Bailer-Jones et al. 2021). Luo et al. (2021) used Gaia DR3 (Gaia Collaboration et al. 2021) for proper motions and distances and used LAMOST spectra for radial velocities ($V_r$; km s$^{-1}$) to obtain the space velocity components U, V, and W, corresponding to the directions of the Galactic center, Galactic rotation, and north Galactic pole, respectively and Galactic orbit parameters.

After rejecting bad spectra, MS stars, WDs, and objects with strong Ca II H&K (λ3933 Å and λ3968 Å), Mg I (λ5183 Å), or Ca II (λ8650 Å) absorption lines. Luo (2021)[3] adopted 1587 hot sd stars by cross-matching Gaia DR2 (Gaia Collaboration et al. 2016) and new spectroscopic surveys LAMOST (Cui et al. 2012). Their spectral and kinematical analysis can be achieved with LAMOST DR7 spectra using their Gaia EDR3 parallaxes and proper motions. Of these hot sd 224 were confirmed for the first time.

According to the He classification scheme (Luo et al. 2021), they identify four groups of He-rich hot subdwarf stars in the $T_{eff}$ – log g and $T_{eff}$ – log($n$He/$n$H) diagrams. Therefore, two extreme He-rich groups (*e*He-1 and *e*He-2) for stars with log($n$He/$n$H) ≥ 0 and two intermediate He-rich groups (*i*He-1 and *i*He-2) for stars with −1 ≤ log($n$He/$n$H) < 0.

---

[1] http://dr7.lamost.org/
[2] ($\mu_\alpha^* = \mu_\alpha \cos \delta$).
[3] https://vizier.cds.unistra.fr/viz-bin/VizieR?-source=J/ApJS/256/28

Both the eccentricity (e) of the orbit and the z-component of the angular momentum ($J_z$) are crucial orbital parameters for understanding the kinematics of hot sd stars. The distribution of the four hot sd He abundance classes and the four groups in the $j_z - e$ diagram is displayed in Figure 10 with (Luo et al. 2021). The two locations identified by Figure (10) of Pauli et al. (2003), region A encompasses thin-disk stars clustering in an area of low eccentricity and $J_z$ of about 1800 kpc km s$^{-1}$. Thick-disk stars with higher eccentricities and lower angular momenta are included in Region B. Halo star candidates are located outside of these two zones, which they refer it as region C. The bulk of stars in the top two panels of their Figure exhibit a continuous distribution from Region A to Region B without any discernible dichotomy, as reported by Luo et al. (2020).

Geier et al. (2022) stated that hot sd hotter than ~ 24,000 K typically exhibits smaller $V_r$ variations and a smaller $V_r$ variability fraction. These few objects could represent yet another subgroup of binaries with longer periods and compact or late-type companions. Herein, and considering this restriction and probabilities (pH ≥ 50%), we have 153 stars within four groups He (i.e., *e*He-1, *e*He-2, *i*He-1, *i*He-2). Table (1) displays their numbers according to our restrictions with their average eccentricities ($\bar{e}$) and z-component of the angular momentum ($\bar{J_z}$).

**Table 1:** The four groups of He with their corresponding numbers, eccentricity, and the z-component of the angular momentum.

| Sample | N | $\bar{e}$ | $\bar{J_z}$ (kpc km s$^{-1}$) |
|---|---|---|---|
| *e*He-1 | 121 | 0.65 | 4288.66 |
| *e*He-2 | 12 | 0.72 | 2289.75 |
| *i*He-1 | 8 | 0.69 | 1008.02 |
| *i*He-2 | 12 | 0.65 | 1530.57 |

Recently, Luo et al. (2024) used the LAMOST DR7 spectra and conducted an analysis of the H, He, C, and N abundances of 210 He-rich hot subdwarf stars collected from various literature references, confirming 191 stars as He-rich hot subdwarf stars. Following their established He classification scheme described in Luo et al. (2021), they identified 127 He-rich stars as *e*He and 64 as *i*He stars.

In what follows, we attention to bold here in our analysis with extreme *e*He-1 rich group (N = 121) only where other groups have low sample numbers and cannot reflect a good feature. Therefore, we reported three programs:
i) The program I was set for inner-halo hot sd with ($T_{eff} \geq$ 24,000 K) and yielded (N = 121) stars.
ii) Program II was set for sdB inner-halo hot sd with (40,000 ≥ $T_{eff}$ ≥ 24,000 K) and yielded (N = 79) stars with $\bar{e}$ = 0.67 and $\bar{J_z}$ = 4288.66 kpc km s$^{-1}$.
iii) Program III was set for sdO inner-halo hot sd with (80,000 ≥ $T_{eff}$ ≥ 40,000 K) and yielded (N = 42) stars with $\bar{e}$ = 0.64 and $\bar{J_z}$ = 2703.63 kpc km s$^{-1}$.

Figure (1) presents the radial velocity ($V_r$; km s$^{-1}$) distribution as a function of Galactic longitude ($l^o$) for all stars observed in sdB and sdO programs, and Figure (2) presents the Galactic spatial velocity components (U, V, W; km s$^{-1}$) along the Galactic coordinates for subsections inner-halo hot sd, sdB, and sdO.

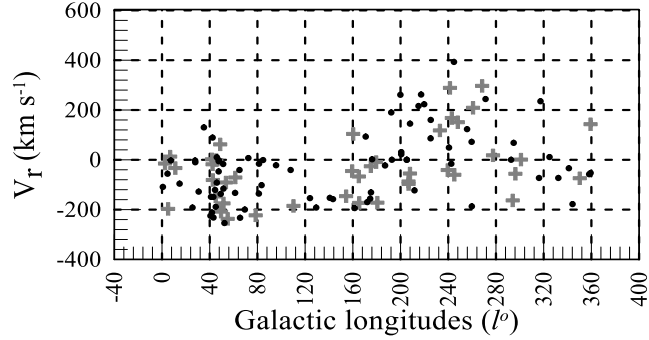

**Fig. 1:** Distribution of ($V_r$; km s$^{-1}$) from LAMOST DR7 spectra versus their Galactic longitudes ($l^o$) for two programs: i) sdB (40,000 ≥ $T_{eff}$ ≥ 24,000 K) (*black closed circles* "79 stars") and ii) sdO (80,000 ≥ $T_{eff}$ ≥ 40,000 K) (*gray closed pluses* "42 stars").

The obtained results are listed in Table (2) with the following format:
Column 1: Program type and its counts (N).
Columns 2 and 3: Coordinate elements, i.e., right ascension and declination (degrees).
Column 4: Effective surface temperature (K).
Columns 5, 6, and 7: Space velocities along the Galactic coordinates ($\overline{U}, \overline{V}, \overline{W}$; km s$^{-1}$).

**Table 2:** Data of programs I, II, and III with Luo (2021).

| Program type | $\alpha_{2000}$ Deg. | $\delta_{2000}$ Deg. | $T_{eff}$ K | U km s$^{-1}$ | V km s$^{-1}$ | W km s$^{-1}$ |
|---|---|---|---|---|---|---|
| **Program I (N = 121)** | | | | | | |
| 1 | 106.56501 | 42.011644 | 32601 | 132 | 130 | -17 |
| 2 | 113.5388 | 52.507966 | 46607 | 149 | -26 | 25 |
| 3 | 113.64929 | 42.650207 | 29489 | 20 | 79 | 55 |
| . | . | . | . | . | . | . |
| . | . | . | . | . | . | . |
| 121 | 273.3217 | 34.31693 | 33067 | -278 | 206 | 98 |
| **Program I (sdB; N = 79)** | | | | | | |
| 1 | 106.565 | 42.01164 | 32601 | 132 | 130 | -17 |
| 2 | 113.6493 | 42.65021 | 29489 | 20 | 79 | 55 |
| 3 | 126.2856 | 48.67533 | 30588 | -56 | 284 | 68 |
| . | . | . | . | . | . | . |
| . | . | . | . | . | . | . |
| 79 | 273.3217 | 34.31693 | 33067 | -278 | 206 | 98 |
| **Program III (sdO; N = 42)** | | | | | | |
| 1 | 113.5388 | 52.50797 | 46607 | 149 | -26 | 25 |
| 2 | 115.0124 | 39.77583 | 79971 | 39 | 162 | 45 |
| 3 | 120.1648 | 58.04277 | 40034 | 67 | 109 | 88 |
| . | . | . | . | . | . | . |
| . | . | . | . | . | . | . |
| 42 | 255.6731 | 31.42189 | 46418 | -17 | 137 | 5 |

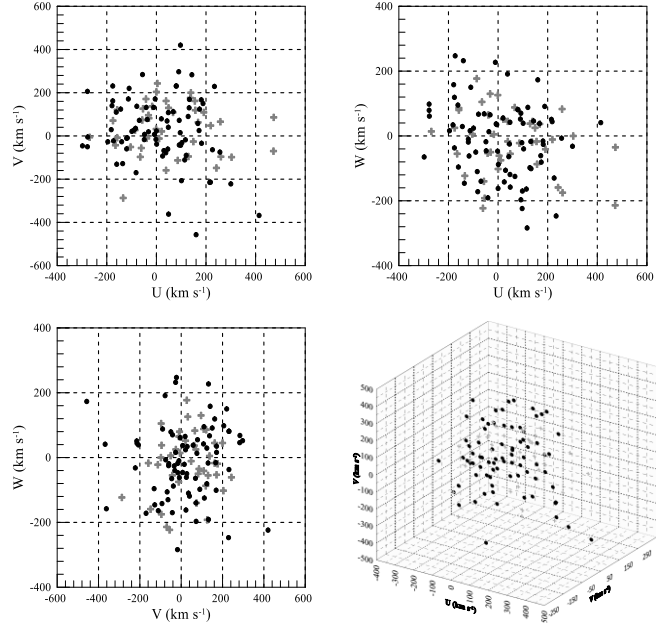

**Fig. 2:** Distribution of the Galactic spatial velocity components (U, V, W; km s$^{-1}$) for sdB *(black closed circles)* and sdO *(gray closed pluses)* programs and their 3D array.

## 3. Semidefinite Programming

Consider $(\dot{x}, \dot{y}, \dot{z})$ as the space velocities of a star located at the Sun. The quadric surface to fit these velocities becomes (Branham 2004)

$$a\dot{x}^2 + b\dot{y}^2 + c\dot{z}^2 + d\dot{x}\dot{y} + e\dot{x}\dot{z} + f\dot{y}\dot{z} + g\dot{x} + h\dot{y} + k\dot{z} + l = 0. \qquad (1)$$

Equation (1) is homogeneous, where *a, …, l* are the ten coefficients that define the quadric that must be determined. The equation may be expressed in more symmetric notation as

$$[\dot{x} \;\; \dot{y} \;\; \dot{z}] \begin{bmatrix} a & d/2 & e/2 \\ d/2 & b & f/2 \\ e/2 & f/2 & c \end{bmatrix} \begin{bmatrix} \dot{x} \\ \dot{y} \\ \dot{z} \end{bmatrix} + [g \;\; h \;\; k] \begin{bmatrix} \dot{x} \\ \dot{y} \\ \dot{z} \end{bmatrix} + l = 0.$$

Use the coefficients of Equation (1), $A = \begin{bmatrix} a & d/2 & e/2 \\ d/2 & b & f/2 \\ e/2 & f/2 & c \end{bmatrix}$ that the matrix be positive-definite and let $v = (\dot{x} \; \dot{y} \; \dot{z})^T$ and $c = (g \; h \; k)^T$.

Then, Equation (1) can be rewritten as
$$(v - c)^T A (v - c) - q^2 = 0, \qquad (2)$$
where
$$q^2 = ag^2 + bh^2 + ck^2 + ekg + dhg + fkh.$$
Define a goodness-of-fit criterion by

$$F = \sum_{i=1}^{m} [(v_i - c)^T A(v_i - c) - q^2]^2, \qquad (3)$$

where *m* denotes the number of data points. Calafiore (2002) referred to this criterion as the "difference of squares" (DOS). It is not a least-squares criterion (Branham 2001) but offers the substantial advantage that with DOS, a unique ellipsoid satisfies the data and represents a global minimum of the criterion. Moreover, convergence is guaranteed if the solution is calculated using semidefinite programming (SDP), which was developed during the 1990s. Although, SDP is an extension of linear programming, developed near the end of the 1940s.

In SDP, one minimizes a linear function subject to the constraint that an affine combination of symmetric matrices is positive semidefinite. Such a constraint is nonlinear and nonsmoothed but convex, so semidefinite programs are convex optimization problems. SDP unifies several standard problems (e.g., linear, and quadratic programming) and finds many applications in engineering and combinational optimization.

Although semidefinite programs are much more general than linear programs, they are not much harder to solve. Most interior-point methods for linear programming have been generalized to semidefinite programs. As in linear programming, these methods exhibit polynomial worst-case complexity and perform very well in practice (Vandenberghe & Boyd 1996).

## 3.1 Solar velocity and components of the velocity ellipsoid

According to Branham (2004), Equation (2) can be converted into Equation (1), and the Solar velocity calculated after the linear terms *g*, *h*, *k* is, by translation of the axes, removed:

$$\begin{bmatrix} \dot{x} \\ \dot{y} \\ \dot{z} \end{bmatrix} = \begin{bmatrix} \dot{x} - l_1 \\ \dot{y} - l_2 \\ \dot{z} - l_3 \end{bmatrix}, \tag{4}$$

where $l_1, l_2, l_3$ are defined to eliminate *g*, *h*, *k* and $[c_1 \quad c_2 \quad c_3] = [l_1 \quad l_2 \quad l_3]$. This is done using the solution of the $3 \times 3$ system.

$$\begin{bmatrix} 2 & d & e \\ d & 2b & f \\ e & f & 2c \end{bmatrix} \cdot \begin{bmatrix} l_1 \\ l_2 \\ l_3 \end{bmatrix} = \begin{bmatrix} -g \\ -h \\ -k \end{bmatrix}. \tag{5}$$

The $l's$ represents the components of the Solar velocity $S_\odot$ found from

$$S_\odot = \sqrt{l_1^2 + l_2^2 + l_3^2}. \tag{6}$$

The translation defined by Equation (2) leaves the coefficients $a, \dots, f$ unchanged, but $l$ becomes $l' = l - al_1^2 - bl_2^2 - cl_3^2 - dl_1l_2 - el_1l_3 - fl_2l_3$. The components of the velocity ellipsoid may be found after an eigenvalue–eigenvector decomposition of the matrix:

$$\begin{bmatrix} 1 & d/2 & e/2 & 0 \\ d/2 & b & f/2 & 0 \\ e/2 & f/2 & c & 0 \\ 0 & 0 & 0 & l' \end{bmatrix}$$

Let $\lambda_1, \lambda_2, \lambda_3, \lambda_4$ be the four eigenvalues. Then, the components of the velocity ellipsoid are found from

$$\begin{bmatrix} \sigma_x \\ \sigma_y \\ \sigma_z \end{bmatrix} = \begin{bmatrix} \sqrt{\lambda_4/\lambda_1} \\ \sqrt{\lambda_4/\lambda_2} \\ \sqrt{\lambda_4/\lambda_3} \end{bmatrix}. \tag{7}$$

The eigenvectors represent the orientations of the velocity ellipsoid.

### 3.2 Kinematical model by matrix elements $\mu_{ij}$

We follow the computational algorithm developed by Elsanhoury et al. (2013) to compute the velocity ellipsoid parameters (VEPs) for the aforementioned three program stars and the Solar elements.

Consider the average space velocity along the Galactic coordinates is $(\overline{U}, \overline{V}, \overline{W};\ \text{km s}^{-1})$:

$$\overline{U} = \frac{1}{N}\sum_{i=1}^{N} U_i\,;\ \overline{V} = \frac{1}{N}\sum_{i=1}^{N} V_i\,;\ \overline{W} = \frac{1}{N}\sum_{i=1}^{N} W_i \tag{8}$$

where $(N)$ denotes the total number of stars in each program. Let $(\xi)$ and its zero point coincides with the center of the distribution. Let $l, m,$ and $n$ be the direction cosines of the axis concerning the shifted one; then, the coordinates $(Q_i)$ of the point $i$, for the $(\xi)$ axis, are as follows:

$$Q_i = l(U_i - \overline{U}) + m(V_i - \overline{V}) + n(W_i - \overline{W}). \tag{9}$$

Consider $(\sigma^2)$ a generalization of the mean square deviation:

$$\sigma^2 = \frac{1}{N}\sum_{i=1}^{N} Q_i^2 \tag{10}$$

From Equations (8), (9), and (10) and after some calculations that

$$\sigma^2 = \underline{x}^T B \underline{x} \tag{11}$$

where $(\underline{x})$ denotes the $(3 \times 1)$ direction cosines of a vector and $(B)$ denotes $(3 \times 3)$ symmetric matrix elements $(\mu_{ij})$;

$$\begin{aligned}
\mu_{11} &= \frac{1}{N}\sum_{i=1}^{N} U_i^2 - (\overline{U})^2;\ \mu_{12} = \frac{1}{N}\sum_{i=1}^{N} U_i V_i - \overline{U}\,\overline{V}; \\
\mu_{13} &= \frac{1}{N}\sum_{i=1}^{N} U_i W_i - \overline{U}\,\overline{W};\ \mu_{22} = \frac{1}{N}\sum_{i=1}^{N} V_i^2 - (\overline{V})^2; \\
\mu_{23} &= \frac{1}{N}\sum_{i=1}^{N} V_i W_i - \overline{V}\,\overline{W};\ \mu_{33} = \frac{1}{N}\sum_{i=1}^{N} W_i^2 - (\overline{W})^2.
\end{aligned} \tag{12}$$

Now, the necessary condition for an extremum is

$$(B - \lambda I)\underline{x} = 0 \tag{13}$$

These are three homogeneous equations in three unknowns with a nontrivial solution if and only if

$$D(\lambda) = |B - \lambda I| = 0, \tag{14}$$

Equation (14) is the characteristic equation for the matrix $(B)$ where $(\lambda)$ is the eigenvalue and $(\underline{x})$ and $(B)$ are given as

$$\underline{x} = \begin{bmatrix} l \\ m \\ m \end{bmatrix} \text{ and } B = \begin{vmatrix} \mu_{11} & \mu_{12} & \mu_{13} \\ \mu_{12} & \mu_{22} & \mu_{23} \\ \mu_{13} & \mu_{23} & \mu_{33} \end{vmatrix}.$$

The required roots (i.e., eigenvalues) are

$$\lambda_1 = 2\rho^{1/3} \cos\frac{\phi}{3} - \frac{k_1}{3};$$

$$\lambda_2 = -\rho^{1/3}\left\{\cos\frac{\phi}{3} + \sqrt{3}\sin\frac{\phi}{3}\right\} - \frac{k_1}{3}; \tag{15}$$

$$\lambda_3 = -\rho^{1/3}\left\{\cos\frac{\phi}{3} - \sqrt{3}\sin\frac{\phi}{3}\right\} - \frac{k_1}{3}.$$

Where

$$\lambda_1 = -(\mu_{11} + \mu_{22} + \mu_{33}),$$
$$\lambda_2 = \mu_{11}\mu_{22} + \mu_{11}\mu_{33} + \mu_{22}\mu_{33} - (\mu_{12}^2 + \mu_{13}^2 + \mu_{22}^2), \tag{16}$$
$$\lambda_3 = \mu_{12}^2\mu_{33} + \mu_{13}^2\mu_{22} + \mu_{23}^2\mu_{11} - \mu_{11}\mu_{22}\mu_{33} - 2\mu_{12}\mu_{13}\mu_{23}.$$

$$q = \frac{1}{3}k_2 - \frac{1}{9}k_1^2; \ r = \frac{1}{6}(k_1 k_2 - 3k_3) - \frac{1}{27}k_1^3. \tag{17}$$

$$\rho = \sqrt{-q^3} \tag{18}$$
$$x = \rho^2 - r^2 \tag{19}$$

and

$$\phi = \tan^{-1}\left(\frac{\sqrt{x}}{r}\right). \tag{20}$$

Depending on the matrix that controls the eigenvalue problem (Equation (13)) for the velocity ellipsoid, we establish analytical expressions of some parameters for the correlation's studies in terms of the matrix elements $(\mu_{ij})$ of the eigenvalue problem for the VEPs.

- **Direction cosine parameters**

The direction cosines $(l_j, m_j, n_j; \forall j = 1,2,3)$ for the eigenvalue problem $(\lambda_j)$, matrix elements $(\mu_{ij})$, and dispersion velocities $(\sigma_j)$ $[i.e., \lambda_j = \sigma_j^2; \forall j = 1,2,3]$ where $(\lambda_1 > \lambda_2 > \lambda_3)$, along three axes (Elsanhoury et al. 2015), are mathematically given as follows:

$$l_j = [\mu_{22}\mu_{33} - \sigma_j^2(\mu_{22} + \mu_{33} - \sigma_j^2) - \mu_{23}^2]/D_j, \tag{21}$$
$$m_j = [\mu_{23}\mu_{13} - \mu_{12}\mu_{33} + \sigma_j^2\mu_{12}]/D_j, \tag{22}$$
$$n_j = [\mu_{12}\mu_{23} - \mu_{13}\mu_{22} + \sigma_j^2\mu_{13}]/D_j, \tag{23}$$

where $(l_j^2 + m_j^2 + n_j^2 = 1)$ is an initial test for our code and $(l_2)$ is known as vertex longitude (Mihalas et al. 1983; Elsanhoury 2016).

$$D_j^2 = (\mu_{22}\mu_{33} - \mu_{23}^2)^2 + (\mu_{23}\mu_{13} - \mu_{12}\mu_{33})^2 + (\mu_{12}\mu_{23} - \mu_{13}\mu_{22})^2 + 2[(\mu_{22} + \mu_{33})(\mu_{23}^2 + \mu_{22}\mu_{33}) + \mu_{12}(\mu_{23}\mu_{13} - \mu_{12}\mu_{33}) + \mu_{13}(\mu_{12}\mu_{23} - \mu_{13}\mu_{22})]\sigma_j^2 + (\mu_{33}^2 + 4\mu_{22}\mu_{33} + \mu_{22}^2 - 2\mu_{23}^2 + \mu_{12}^2 + \mu_{13}^2)\sigma_j^4 - 2(\mu_{22} + \mu_{33})\sigma_j^6 + \sigma_j^8.$$

- **Galactic longitude and latitude parameters**

  Let $(L_j$ and $B_j; \forall j = 1,2,3)$ be the Galactic longitude and latitude of the directions, respectively, which correspond to the extreme values of the dispersion; then,

$$L_j = tan^{-1}\left(\frac{-m_j}{l_j}\right), \tag{24}$$

$$B_j = sin^{-1}(-n_j), \tag{25}$$

- **Solar elements**

  For our program star having space velocities $(\bar{U}, \bar{V}, \bar{W})$, the components of the Sun's velocities are referred to as $(U_\odot, V_\odot,$ and $W_\odot)$, where $(U_\odot = -\bar{U})$, $(V_\odot = -\bar{V})$, and $(W_\odot = -\bar{W})$. Therefore, the Solar elements $(S_\odot, l_A, b_A)$ with spatial velocity considered may take the following:

$$S_\odot = \sqrt{\bar{U}^2 + \bar{V}^2 + \bar{W}^2}, \tag{26}$$

$$l_A = tan^{-1}\left(\frac{-\bar{V}}{\bar{U}}\right), \tag{27}$$

$$b_A = sin^{-1}\left(\frac{-\bar{W}}{S_\odot}\right). \tag{28}$$

where $(S_\odot)$ denotes the absolute value of the Sun's velocity relative to our three program stars under consideration, $(l_A)$ is the Galactic longitude, and $(b_A)$ is the Galactic latitude of the Solar apex.

## 4. Galactic Rotation Kinematics

The first branch heavily depends on assumptions about the form of the Galactic potential and the distance scale (see Majewski et al. 2006). Other issues include the inability of tidal streams to distinguish between stellar orbits (Eyre & Binney 2009; Schönrich 2012). The change in velocity ellipsoid concerning height above the Galactic plane and metallicity has also been investigated (i.e., thin and thick disks); the two components differ in their spatial distribution, metallicity, and kinematics. In the spatial distribution, the range of scale height and length for the thin disk is about 200–369 pc and 1.00–3.70 kpc, respectively, whereas the thick disk has a scale height of 600–1000 pc and a scale length of 2.00–5.50 kpc (Du et al. 2003, 2006; Bilir et al. 2006, 2008; Karaali et al. 2007; Jurić et al. 2008; Yaz & Karaali 2010; Chang et al. 2011; Jia et al. 2014; Chen et al. 2017; Wan et al. 2017).

The main source of information used to derive Galactic parameters, such as linear rotation velocity $(V_o;$ km s$^{-1})$ (Mihalas & Binney 1981) of the Galaxy near-circular orbit of radius $(R_o = 8.20 \pm 0.10$ kpc) (Bland-Hawthorn et al. 2019), e.g., by the application of the Oort constants (A, B; km s$^{-1}$kpc$^{-1}$) (Oort 1927a & 1927b) with few more attempts to use luminous stars (e.g., Burton & Bania 1974). Most of the recent research on $R_o$ and $V_o$ focused on modeling streams in the Galactic halo (Ibata et al. 2001; Majewski et al. 2006), from microwave amplification by stimulated emission of radiation (Reid & Brunthaler 2004), the Galactic center, molecular clouds, and radio studies of the H I terminal velocity (McMillan 2011).

The disk of our Galaxy is considered to be axisymmetric (Oort 1927b), and its stellar population is in a state of differential rotation (i.e., random motion) around an axis through the Galactic center, undergoing stars' nearly circular orbits about the Galactic center. The first-ever proof of the presence of the differential Galactic rotation was provided by Oort (1927a, 1927b) at a $V_o$ of 247.5 km s$^{-1}$. Mihalas & Binney (1981) obtained ($200 \leq V_o \leq 300$ km s$^{-1}$), with the "best" result being about ($V_o = 250$ km s$^{-1}$), which depended on the radial velocities of globular clusters or spheroidal component stars in our Galaxy or external Galaxies in the local group. Bovy (2017) and Li et al. (2019) obtained $V_o$ of about 204.00 and 213.70 km s$^{-1}$, respectively. Recently, we obtained $V_o$ of 221.20 (Nouh & Elsanhoury 2020), 256.64 ± 0.37 (Elsanhoury et al. 2021), and 213.81 ± 14.61 km s$^{-1}$ (Elsanhoury & Al-Johani 2023).

The angular rotation rate ($\omega_o$) of the Galaxy at a radius ($R_o$) is given by $V_o = \omega_o R_o = |A - B| R_o$. In what follows and depending on our constructed model of the inner-halo hot sd, we can compute Oort's constants (A & B) kinematically.

The major quantity we can devote to the kinematics of hot sd is the association between the ratio of ($\sigma_2/\sigma_1$) and Oort's constants (A & B), i.e.,

$$\left(\frac{\sigma_2}{\sigma_1}\right)^2 = \frac{-B}{A - B},$$

or

$$\frac{-B}{A} = \frac{1}{(\sigma_1/\sigma_2)^2 - 1}. \tag{29}$$

Considering that the average numerical value of the angular rotation rate ($|A - B|$; km s$^{-1}$kpc$^{-1}$) is about 26.07 ± 5.10 based adopted with our computations (Elsanhoury & Al-Johani 2023).

We herein developed our code to serve computing VEPs, including the ratio ($\sigma_2/\sigma_1$) devoted to our three program stars; the obtained results ranged from 0.69 to 0.86. The original numerical results are listed in Table (3) and Table (4), including the Solar elements.

**Table 3:** Our kinematical parameters of programs I, II, and III retrieved with inner-halo hot sd ($d \leq 15$ kpc) at high Galactic latitudes ($b^o \geq 20$).

| Parameters | Program I: 121 points (inner-halo hot sd) ($T_{eff} \geq 24,000$) | Program II: 79 points (sdB inner-halo hot sd) ($40,000 \geq T_{eff} \geq 24,000$) | Program III: 42 points (sdO inner-halo hot sd) ($80,000 \geq T_{eff} \geq 40,000$) |
|---|---|---|---|
| ($\bar{U}, \bar{V}, \bar{W}$; km s$^{-1}$) | 25.73 ± 5.07, 28.79 ± 5.37, −14.51 ± 3.81 | 12.56 ± 3.54., 25.62 ± 5.06, −13.43 ± 3.66 | 50.50 ± 7.11, 34.74 ± 5.89, −16.52 ± 4.06 |
| ($\lambda_1, \lambda_2, \lambda_3$; km s$^{-1}$) | 26225.50, 19685.90, 10316.00 | 27426.30, 19577.40, 11209.20 | 30113.00, 14445.50, 7855.45 |
| ($\sigma_1, \sigma_2, \sigma_3$; km s$^{-1}$) | 161.94 ± 12.73, 140.31 ± 11.85, 101.57 ± 10.08 | 165.61 ± 12.87, 139.92 ± 11.83, 105.87 ± 10.29 | 173.53 ± 13.17, 120.19 ± 10.96, 88.63 ± 9.41 |
| ($\bar{\sigma_o}$; km s$^{-1}$) | 237.12 ± 15.40 | 241.27 ± 15.53 | 228.94 ± 15.13 |
| ($l_1, m_1, n_1$)° | 0.8965, -0.3606, -0.2572 | 0.6388, -0.7491, -0.1752 | 0.9760, 0.0771, -0.2037 |
| ($l_2, m_2, n_2$)° | -0.3454, -0.9327, 0.1037 | -0.6735, -0.6546, 0.3434 | -0.0264, 0.9703, 0.2405 |
| ($l_3, m_3, n_3$)° | 0.2773, 0.0041, 0.9608 | 0.3719, 0.1014, 0.9227 | 0.2162, -0.2294, 0.9490 |
| $L_j$, j = 1,2,3 | 21.92, 110.32, 179.16 | 49.54, 135.81, 164.75 | -4.52, -91.56, -133.30 |
| $B_j$, j = 1,2,3 | -14.91, 5.95, 73.90 | -10.09, 20.08, 67.32 | -11.75, 13.92, 71.63 |
| ($S_\odot$; km s$^{-1}$) | 41.24 ± 6.42 | 31.53 ± 5.62 | 63.48 ± 7.97 |
| ($l_A, b_A$)° | −48.21, 20.59 | -63.89, 25.21 | −34.52, 15.09 |

## 5. Results

Based on the aforementioned models, we developed our Mathematica routine to compute the kinematics and VEPs of program stars I (121 points), II (79 points), and III (42 points). Table (3) presents the results obtained, including the following: spatial velocities $(\bar{U}, \bar{V}, \bar{W};$ km s$^{-1}) = (25.73 \pm 5.07, 28.79 \pm 5.37, -14.51 \pm 3.81$ "Program I"; $12.56 \pm 3.54., 25.62 \pm 5.06, -13.43 \pm 3.66$ "Program II"; $50.50 \pm 7.11, 34.74 \pm 5.89, -16.52 \pm 4.06$ "Program III"), dispersion velocities $(\sigma_1, \sigma_2, \sigma_3;$ km s$^{-1}) = (161.94 \pm 12.73, 140.31 \pm 11.85, 101.57 \pm 10.08$ "Program I"; $165.61 \pm 12.87, 139.92 \pm 11.83, 105.87 \pm 10.29$ "Program II"; $173.53 \pm 13.17, 120.19 \pm 10.96, 88.63 \pm 9.41$ "Program III"), direction cosines $(l_j, m_j, n_j; \forall j = 1,2,3)$, Solar velocity $(S_\odot;$ km s$^{-1}) = (41.24 \pm 6.42$ "Program I"; $31.53 \pm 5.62$ "Program II"; $63.48 \pm 7.97$ "Program III"), and others. As can be seen from Table (4), our ratios of $(\sigma_2/\sigma_1)$ are about (0.86 "Program I"; 0.84 "Program II"; 0.69 "Program III"); similarly, we obtained $(\sigma_3/\sigma_1)$ as (0.62 "Program I"; 0.64 "Program II"; 0.51 "Program III").

In Table (5), we listed the Oort constant adopted by some earlier studies. Elsanhoury and Al-Johani (2023) used two samples of K dwarf stars based on Gaia DR3 data within Solar neighborhood ≤ 25 pc with b ≥ 20o and b < 20o respectively. Wang et al. (2021) derived Oort constants using a sample of 5,627 A-type stars within 0.60 kpc selected from the LAMOST surveys. Nouh and Elsanhoury (2020) calculated the Oort constants using a sample of halo red giants based on the space and radial velocities of 1,583 red giants' stars collected from SEGUE-1 and SEGUE-2 surveys. Krisanova et al. (2020) computed Oort constants by studying a sample of more than 25,000 young stars with proper motions and trigonometric parallaxes from the Gaia DR2 catalog. Bobylev and Bajkova (2019) studied the kinematics of hot dim stars from the catalog by Geier et al. (2019) with 39,800 hot sd candidates selected by them from the Gaia DR2 catalog using data from several multiband photometric sky surveys.

Comparison of the results from the present analysis with those listed in Table (5) showed good agreement for the first Oort constant A, but we found differences for the second Oort constant B. The different values of the Oort constants, i.e. A and B can be attributed to different calculation methods and large distances of faint stars from the Sun, as Hanson (1987) emphasised that the mean distances of these "faint anonymous stars" are of the order of 1 kpc or less.

**Table 4:** Velocity dispersions and Solar velocities for our inner-halo hot sd and other components of the disks with different authors.

| Type | $\sigma_1$ | $\sigma_2$ | $\sigma_3$ | $S_\odot$ | $(\sigma_2/\sigma_1)$ | $(\sigma_3/\sigma_1)$ | Reference |
|---|---|---|---|---|---|---|---|
| Program I – 121 points | 161.94 ± 12.73 | 140.31 ± 11.85 | 101.57 ± 10.08 | 41.24 ± 6.42 | 0.86 | 0.62 | Current work |
| Program II (sdB; 79 points) | 165.61 ± 12.87 | 139.92 ± 11.83 | 105.87 ± 10.29 | 31.53 ± 5.62 | 0.84 | 0.64 | Current work |
| Program III (sdO; 42 points) | 173.53 ± 13.17 | 120.19 ± 10.96 | 88.63 ± 9.41 | 63.48 ± 7.97 | 0.69 | 0.51 | Current work |
| Inner halo ($d \leq 15$ kpc) | 265.86 | 187.48 | 138.29 | 213.36 ± 14.61 | 0.70 | 0.52 | Nouh & Elsanhoury (2020) |
| Hot sd ($b^o \geq 20$) | 45.23 ± 0.86 | 37.68 ± 1.80 | 30.10 ± 0.84 | - | 0.83 | 0.67 | Bobylev and Bajkova (2019) |
| Halo disk | 160 | 90 | 90 | - | 0.56 | 0.56 | Yan et al. (2019) |
| Thin disk | 35 | 20 | 16 | - | 0.57 | 0.46 | Yan et al. (2019) |
| Thick disk | 67 | 38 | 35 | - | 0.57 | 0.52 | Yan et al. (2019) |
| Hot WD | 29.10 | 17.50 | 23.31 | 11.28 | 0.60 | 0.80 | Elsanhoury et al. (2015) |
| Thin disk | 63 ± 6.0 | 39 ± 4.0 | 39 ± 4.0 | - | 0.62 | 0.62 | Soubiran (2003) |
| Thick disk | 39 ± 2.0 | 20 ± 2.0 | 20 ± 1.0 | - | 0.51 | 0.51 | Soubiran (2003) |
| (Oe5–B5), brighter than 6.00 mag. | 11.00 | 9.00 | 5.40 | - | 0.82 | 0.49 | Nordstrom (1936) |
| (B8–B9), brighter than 6.00 mag. | 12.30 | 10.20 | 9.10 | - | 0.83 | 0.74 | Nordstrom (1936) |

**Table 5:** The rotation constates with our three program stars and recently computed ones.

| $(\sigma_2/\sigma_1)$ | (A; km s$^{-1}$ kpc$^{-1}$) | (B; km s$^{-1}$ kpc$^{-1}$) | Reference |
|---|---|---|---|
| 0.86 | 6.79 ± 0.38 | −19.28 ± 0.23 | Current work |
| 0.84 | 7.68 ± 0.36 | −18.39 ± 0.24 | Current work |
| 0.69 | 13.66 ± 0.27 | −12.41 ± 0.28 | Current work |
| 0.72 | 12.91 ± 0.28 | −13.16 ± 0.28 | Elsanhoury & Al-Johani (2023) |
| - | 16.31 ± 0.61 | −11.99 ± 0.79 | Wang et al. (2021) |
| 0.71 | 15.60 ± 0.03 | −13.90 ± 1.8 | Nouh & Elsanhoury (2020) |
| - | 15.73 ± 0.32 | −12.67 ± 0.34 | Krisanova et al. (2020) |
| 0.83 | 9.41 ± 0.92 | -14.24 ± 1.38 | Bobylev & Bajkova (2019) |

## 6. Discussion and Conclusions

This study investigated inner-halo hot sd stars, with effective surface temperatures of ($T_{eff} \geq 24{,}000$ K). Their subsections, namely, sdB and sdO, had effective surface temperatures of ($40{,}000 \geq T_{eff} \geq 24{,}000$ K) and ($80{,}000 \geq T_{eff} \geq 40{,}000$ K) respectively.

First, we downloaded raw data retrieved by Gaia DR2 (Gaia Collaboration et al. 2016) which improved with used Gaia DR3 (Gaia Collaboration et al. 2021) for proper motions and distances to obtain the space velocity components ($U, V, W$; km s$^{-1}$) and LAMOST DR7 (Cui et al. 2012) pointed out with ($d \leq 15$ kpc), located at high Galactic latitudes ($b^o \geq 20$), with probability adopted with the data source (pH $\geq 50\%$). Therefore, we have three programs: i) inner-halo hot sd (121 stars; Program I), ii) inner-halo hot sd "sdB" (79; Program II), and iii) inner-halo hot sd "sdO" (42; Program III).

Table (4) presents the dispersion velocities, Solar velocities, and their ratios, i.e., ($\sigma_1, \sigma_2, \sigma_3$), $S_\odot$, ($\sigma_2/\sigma_1$), and ($\sigma_3/\sigma_1$), respectively. Our average numerical results were ($\overline{\sigma_1}, \overline{\sigma_2}, \overline{\sigma_3}$; km s$^{-1}$) = (167.03 ± 12.92, 133.47 ± 11.55, 98.69 ± 9.93), which indicated a good agreement with a range of 138 – 266 for the inner halo located at ($d \leq 15$ kpc) adopted with Nouh and Elsanhoury (2020). The ratios of $\overline{(\sigma_2/\sigma_1 = 0.79)}$ and $\overline{(\sigma_3/\sigma_1 = 0.59)}$ were consistent with the other obtained ones, ($0.51 \leq \sigma_2/\sigma_1 \leq 0.83$) and ($0.46 \leq \sigma_2/\sigma_1 \leq 0.80$), respectively.

Adopting our angular rotation rate $|A - B| \sim 26.07 \pm 5.10$ km s$^{-1}$ kpc$^{-1}$ (Elsanhoury and Al-Johani, 2023) and the dynamical association between (A & B) as $(\sigma_2/\sigma_1)^2 = -B/(A - B)$, we may estimate Oort's constants as ($A$ & $B$; km s$^{-1}$kpc$^{-1}$) = (6.79 ± 0.38 & −19.28 ± 0.23 "Program I"; 7.68 ± 0.36 & −18.39 ± 0.24 "Program II"; 13.66 ± 0.27 & −12.41 ± 0.28 "Program III").

The following conclusions were obtained:
- With eccentricity (e = 0.65) and the z-component of the angular momentum ($J_z$ = 4288.66 kpc km s$^{-1}$) for one of the extreme He-rich groups (*e*He-1), we devoted three sets of program stars were considered: Program I ($T_{eff} \geq 24{,}000$ K), Program II ($40{,}000 \geq T_{eff} \geq 24{,}000$ K), and Program III ($80{,}000 \geq T_{eff} \geq 40{,}000$ K) with 121, 79, and 42 points, respectively.
- We developed our Mathematica code to estimate their inner kinematics and VEPs, e.g., total spatial velocities along the Galactic coordinates $\left(\sqrt{\overline{U}^2 + \overline{V}^2 + \overline{W}^2}; \text{km s}^{-1}\right)$ = (41.25 ± 6.42 "Program I"; 31.54 ± 5.62 "Program II"; 63.48 ± 7.97 "Program III"), dispersion velocities $\left(\overline{\sigma_o} = \sqrt{\sigma_1^2 + \sigma_2^2 + \sigma_3^2}; \text{km s}^{-1}\right)$ = (237.12 ± 15.40 "Program I"; 241.27 ± 15.53 "Program II"; 228.94 ± 15.13 "Program III"), direction cosines $(l_j, m_j, n_j; \forall j = 1,2,3)$ with negative values of the vertex longitudes ($l_2$) as mentioned in our previous investigations (i.e., $l_2^o$) = (−0.3454 "Program I"; −0.6735 "Program II"; −0.0264 "Program III") as e.g. $-0°.006394$ and $-0°.630221$ (Elsanhoury

and Al-Johani 2023); $-0°.5410, -0°.4937$, and $-0°.9495$ (Elsanhoury et al. 2021); $-0°.92, -0°.906$, and $-0°.697$ (Nouh and Elsanhoury 2020), and the Solar velocities $(\overline{S_\odot}; \text{km s}^{-1}) = (45.42 \pm 6.74)$.

- The determination of the average local Galactic rotational constants, i.e., Oort's constants, are $(A \& B; \text{km s}^{-1}\text{kpc}^{-1}) = (9.38 \pm 0.33, -16.69 \pm 0.25)$, these obtained values well agreed with other studies up to 67% and 79% respectively with other adopted authors.

- Lewis (1990) and Hanson (1987) concluded that the large distances and/or the choice of the kinematical model analysis may result in the under-and/or overestimation of Oort's constants.

We conclude that the cross-match between Gaia DR3 and LAMOST DR7 has improved the overall determined parameters of the inner halo sd, especially those obtained from the astrometric observations, by inferring their uncertainty. This would lead to more accurate properties. However, further studies on these types are needed to confirm the current results and support our conclusions.

**Acknowledgments**

The author is deeply thankful to the referee for his/her valuable and constructive comments that improved the original manuscript. This work presents results from the European Space Agency space mission Gaia. Gaia data are being processed by the Gaia Data Processing and Analysis Consortium (DPAC). Funding for the DPAC is provided by national institutions, in particular, the institutions participating in the Gaia MultiLateral Agreement (MLA). The Gaia mission website is https://www.cosmos.esa.int/gaia. The Gaia archive website is https://archives.esac.esa.int/gaia.

Guoshoujing Telescope (Large Sky Area Multi-Object Fiber Spectroscopic Telescope LAMOST) is a National Major Scientific Project built by the Chinese Academy of Sciences. Funding for the project has been provided by the National Development and Reform Commission. LAMOST is operated and managed by the National Astronomical Observatories, Chinese Academy of Sciences.

**Data availability**

Datasets analyzed during the current study were derived from the following public domain resources: "Gaia Collaboration, VizieR On-line Data Catalog: J/ApJS/256/28, October 2021, originally published. in: The Astrophysical Journal Supplement Series" https://vizier.cds.unistra.fr/viz-bin/VizieR?-source=J/ApJS/256/28.

**ORCID iDs**

W. H. Elsanhoury  https://orcid.org/0000-0002-2298-4026

## References

- Bailer-Jones C. A. L., Rybizki J., Fouesneau M., et al., 2021, Astron. J., 161, 147
- Bilir S., Cabrera-Lavers A., Karaali S., et al., 2008, Pub. Astron. Soc. Austr., 25, 69
- Bilir S., Karaali S., and Gilmore G., 2006, Mon. Not. Roy. Astron. Soc., 366, 1295
- Bland-Hawthorn J., Sharma, S., et al., 2019, Mon. Not. Roy. Astron. Soc., 486, 1167
- Bobylev V. V. and Bajkova A. T., 2019, Astron. Lett., 45, 580
- Bovy J., 2017, Mon. Not. Roy. Astron. Soc., 468, L63
- Branham Jr. R. L., 2001, New Astron. Rev., 45, 649
- Branham Jr. R. L., 2004, Astron. & Astrophys., 421, 977
- Burton W. B., Bania T. M., 1974, Astron. & Astrophys., 33, 425
- Calafiore G., 2002 IEEE Trans. Sys., Man & Cyber., Part A, 32, 269
- Carney R. and Latham D., 1986, Astrophys. J., 92, 60
- Chang C. K., Ko C. M., and Peng T. H., 2011, Astrophys. J., 740, 34
- Charpinet S., Green E. M., Baglin A., et al., 2010, Astron. & Astrophys., 516, 6
- Chen B. Q., Liu X. W., Yuan H. B., et al., 2017, Mon. Not. Roy. Astron. Soc., 464, 2545
- Cui X.-Q., Zhao Y.-H., Chu Y.-Q., et al., 2012, Res. Astron. Astrophys., 12, 1197
- Culpan R., Geier S., Reindl N., Pelisoli I., Gentile Fusillo N., and Vorontseva A., 2022, Astron. & Astrophys., 662, A40
- Deason A. J., Belokurov V., Evans N. W., 2011, Mon. Not. Roy. Astron. Soc., 416, 2903
- Drilling J. S., Jeffery C. S., Heber U., Moehler S., and Napiwotzki R., 2013, Astron. Astrophys.,551, A31
- Du C. H., Ma J., Wu Z. Y., and Zhou X., 2006, Mon. Not. Roy. Astron. Soc., 372, 1304
- Du C. H., Zhou X., Ma J., et al., 2003, Astron. & Astrophys., 407, 541
- Elsanhoury W. H. and Al-Johani A. S., 2023, Astron. Nachr., 344, issue 7
- Elsanhoury W. H., 2016, Astrophys., 59, pp. 246
- Elsanhoury W. H., Nouh M. I. and Abdel-Rahman H. I., 2015, Rev. Mex. Astron. Astrofís., Vol. 51, 197
- Elsanhoury W. H., Nouh M. I., Branham Jr. R. L., and Al-Johani A. S., 2021, Astron. Nachr., 342 (989), 989
- Elsanhoury W. H., Sharaf M. A., Nouh M. I., et al., 2013, The Open Astron. J., 6, 1
- Eyre A. and Binney J., 2009, Mon. Not. Roy. Astron. Soc., 400, 548
- Gaia Collaboration, Brown A. G. A., Vallenari A., et al., 2018, Astron. & Astrophys., 616, A1
- Gaia Collaboration, Brown A. G. A., Vallenari A., et al., 2021, Astron. & Astrophys., 649, A1
- Gaia Collaboration, Prusti T., de Bruijne J. H. J., et al., 2016, Astron. & Astrophys., 595, A1
- Gaia Collaboration, Vallenari A., and Brown A. G. A., 2023, Astron. & Astrophys., 674, A1
- Geier R. R., Gentile Fusillo, N. P., and Marsh, T. R., 2019, Astron. Astrophys., 621, 38

- Geier S., 2015, Astrono. Nachr., Vol. 336, Issue 5, p.437
- Geier S., Dorsch M., Pelisoli I., Reindl N., et al., 2022, Astron. Astrophys., 661, 113
- Geier S., Heber U., Heuser C., et al., 2013, Astron. & Astrophys., 551, L4
- Gilmore G., and Reid N., 1983, Mon. Not. Roy. Astron. Soc., 202, 1025
- Greenstein J. L., 1954, Astro. J., 59, 322
- Hanson R. B., 1987, Astron. Nachr., 94, 409
- Heber U., 1986, Astron. & Astrophys., 155, 33
- Heber U., 2009, Ann. Rev. Astron. Astrophys., 47, 211
- Heber U., 2009, Ann. Rev. Astron. Astrophys., 47, 211
- Heber U., 2016, Pub. Astron. Soc. Pas., 128, 082001
- Ibata R., Lewis G. F., Irwin M., Totten E., Quinn T., 2001, Astrophys. J., 551, 294
- Jia Y. P., Du C. H., et al., 2014, Mon. Not. Roy. Astron. Soc., 441, 503
- Jurić M., Ivezić Ž, Brooks A., et al., 2008, Astrophys. J., 673, 864
- Karaali S., Bilir S., Yaz E., Hamzaoğlu E., and Buser, R., 2007, Pub. Astron. Soc. Austr., 24, 208
- Krisanova O. I., Bobylev V. V., and Bajkova A. T., 2020, Astron. Lett., 46(6), 370
- Lei Z., Zhao J., Németh P., et al., 2020, Astrophys. J., 889, 117
- Lewis J. R., 1990, Mon. Not. Roy. Astron. Soc., 244, 247-253
- Li C., Zhao G., and Yang C., 2019, Astrophys. J., 872(2), 205
- Liu J.-C., Zhu Z., and Hu B., 2011, Astron. Astrophys. 536, A102
- Luo A., Zhang J., Chen J., et al., 2014, in IAU Symp. 298, Setting the Scene for Gaia and LAMOST (Cambridge: Cambridge Univ. Press), 428
- Luo A.-L., Zhang H.-T. Zhao Y.-H., et al., 2012, Res. Astron. Astrophys., 12, 1243
- Luo Y., Németh P., Wang K., Pan Y., 2024, Astrophys. J. Suup., 271, Issue 1, 21
- Luo Y., Németh P., Wang K., Wang X., Han Z., 2021, Astrophys. J. Suup., 256, Issue 2, id.28, 22
- Majewski S. R., Law D. R., Polak A. A., Patterson R. J., 2006, Astrophys. J., 637, L25
- Martin P., Jeffery C. S., Naslim N., Woolf V. M., 2017, Mon. Not. Roy. Astron. Soc., 467, Issue 1, pp. 68-82
- Maxted P. F. L., Marsh T. R., and North R. C., 2000, Mon. Not. Roy. Astron. Soc., 317, L41
- McCarthy I. G., Font A. S., Crain R. A. Deason A. J., Schaye J. Theuns T., 2012, Mon. Not. Roy. Astron. Soc., 420, Issue 3, pp. 2245-2262
- McMillan P., 2011, Mon. Not. Roy. Astron. Soc., 414, 2446
- Mihalas D., Binney J., and Antonov V. A., 1983, Astrofizika, 19, 505
- Mihalas, D. and Binney J., 1981, Galactic Astronomy (San Francisco, CA: Freeman)
- Moehler S., Richtler T., de Boer K. S., Dettmar R. J., and Heber U., 1990, Astron. & Astrophys., 86, 53
- Nordstrom H., Meddelanden fran Lunds Astronomiska Observatorium, 1936, Series II, Vol. 79, p.3-196
- Nouh M. I. and Elsanhoury, W. H., 2020, Astrophys., Vol. 36, No. 2, pp. 179-189
- O'Connell R. W. 1999, Ann. Rev. Astron. Astrophys., 37, 603

- Oort J. H., 1927a, Bull. Astron. Inst. Netherl., 3, 275
- Oort J. H., 1927b, Bull. Astron. Inst. Netherl., 4, 91
- Pauli E. -M., Napiwotzki R., Altmann M., et al., 2003, Astron. Astrophys., 400, 877
- Reid M. J., Brunthaler A., 2004, Astrophys. J., 616, 872
- Sargent W. L. W. and Searle L., 1968, Astrophys. J. 152, 443
- Schönrich R., 2012, Mon. Not. Roy. Astron. Soc., 427, 274
- Silvotti R., Schuh S., Janulis R., et al., 2007, Nature, 449, 189
- Sion E. M., Greenstein J. L., Landstreet J., Liebert J., Shipman H. L., and Wegner G., 1983, Astrophys. J., 269, 253
- Soubiran C., Bienaume O., and Siebert A., 2003, Astron. & Astrophys., 398, 141
- Vandenberghe L., Boyd S., 1996, SIAM Rev., 38, 49
- Wan J. C., Liu C., and Deng L. C., 2017, Res. Astron. Astrophys., 17, 79
- Wang F., Zhang H. W., Huang Y., Chen B. Q., Wang H. F., and Wang C., 2021, Mon. Not. Roy. Astron. Soc., 504 (1), 199
- Yan Y., Du C., Liu S., Li H., Shi J., Chen Y., Ma J., and Wu Z., 2019, Astrophys. J., 880, 36
- Yaz E. and Karaali S., 2010, New Astron., 15, 234